\documentclass[runningheads]{llncs}
\usepackage{graphicx}
\usepackage[figurename=Fig.]{caption}
\usepackage{subcaption}
\usepackage{wrapfig}
\usepackage{marvosym}
\usepackage[colorlinks = true,
            linkcolor = blue,
            urlcolor  = blue,
            citecolor = blue,
            anchorcolor = blue]{hyperref}

\begin{document}
\title{Structural Similarity based Anatomical and Functional Brain Imaging Fusion}
%
%
\author{Nishant Kumar\inst{1}\textsuperscript{(\Letter)} \and
Nico Hoffmann\inst{2} \and
Martin Oelschl$\ddot{a}$gel\inst{3} \and Edmund Koch\inst{3} \and Matthias Kirsch\inst{4} \and Stefan Gumhold\inst{1}}
\authorrunning{N. Kumar et al.}
%
\institute{Computer Graphics and Visualisation, Technische Universit$\ddot{a}$t Dresden, \\ 01062 Dresden, Germany \\
\email{nishant.kumar@tu-dresden.de}
\and Institute of Radiation Physics, Helmholtz-Zentrum Dresden-Rossendorf,\\ 01328 Dresden, Germany \and 
Clinical Sensoring and Monitoring, Technische Universit$\ddot{a}$t Dresden, \\ 01307 Dresden, Germany \and Department of Neurosurgery, University Hospital Carl Gustav Carus, \\ 01307 Dresden, Germany}
\maketitle 
\begin{abstract}
Multimodal medical image fusion helps in combining contrasting features from two or more input imaging modalities to represent fused information in a single image. One of the pivotal clinical applications of medical image fusion is the merging of anatomical and functional modalities for fast diagnosis of malign tissues. In this paper, we present a novel end-to-end unsupervised learning based Convolutional neural network (CNN) for fusing the high and low frequency components of MRI-PET grayscale image pairs publicly available at ADNI by exploiting Structural Similarity Index (SSIM) as the loss function during training. We then apply color coding for the visualization of the fused image by quantifying the contribution of each input image in terms of the partial derivatives of the fused image. We find that our fusion and visualization approach results in better visual perception of the fused image, while also comparing favorably to previous methods when applying various quantitative assessment metrics.  

\end{abstract}

\section{Introduction}
A rapid advancement in sensor technology has improved medical prognosis,
surgical navigation and treatment. For example, anatomical
modalities such as Magnetic resonance imaging (MRI) and Computed Tomography (CT) reveals the structural
information of the brain like the location of tumor as well as white and gray matter while modalities such as Positron emission tomography
(PET) provides functional information like glucose metabolism. The hybrid blend of PET-CT acquisition hardware provides fast and accurate attenuation correction and helps in combining anatomical and functional information. However it exposes patients to high level of X-Ray and ionizing radiation. The integrated MRI-PET scanners results in high tissue contrast with significantly low radiation dose. But the development of a robust hybrid MRI-PET hardware is challenging due to compatibility issue of PET detectors in a high magnetic field environment of MRI. The post-hoc fusion of MRI-PET image pairs overcomes the challenges of fully integrated MRI-PET scanners and helps medical personnel to better diagnose brain abnormalities such as glioma and Alzheimer's disease \cite{ref1,ref_1}.

Most of the past image fusion methods proposed a three
step approach to the fusion problem. First, the source images were transformed into a particular domain using approaches such as multi-scale decomposition \cite{ref2,ref3,ref4,ref5,ref6}, sparse representation \cite{ref7,ref8}, mixture of multi-scale decomposition and sparse representation \cite{ref9} and Intensity-Hue-Saturation \cite{ref10} among others. Then,
the transformed coefficients are combined using a predefined
coefficient grouping based fusion strategy such as max selection and weighted-averaging. Finally, the fused image is reconstructed
by taking the inverse of the transformation strategy. However, the intricacy of these methods leads to the computational inefficiency making them unrealistic for the real time setup \cite{ref11}. CNN based medical image fusion \cite{ref_proc15} has been actively studied in the past. However, these methods train the network on natural images due to the unavailability of large preregistered medical image pairs. The acquisition method of natural images differ from PET images since PET accumulates nuclear tracers depending on positron range, photon collinearity or the width of the detector element that results in a smooth low resolution acquisition without clear interfaces between certain tissues. The high resolution MRI such as T1-MPRAGE on the other hand are acquired in spatial frequency domain by varying the sequence of RF pulses. Hence, the aspects of human visual system that are tuned to process natural images are not equally useful for MRI-PET images due to which the selection of a proper
objective assessment metric is challenging \cite{ref12}. Secondly, there are no ground truth in a fusion problem due to which proper selection of the loss layer becomes critical.

Therefore, we propose a fast grayscale anatomical and functional medical image fusion approach in an end-to-end unsupervised learning network trained on publicly available medical image pairs. Additionally, the fusion result is visualized based on the contribution of the input images to the fused output image. The computational efficiency of our combined fusion and visualisation framework has the potential of real time clinical application in future.


\section{Methods}
\subsection{Fusion framework}
The fusion architecture in Fig.1. takes two grayscale input images $I_1$ and $I_2$ and generates a grayscale fused image $I_F$. The network consist of three different strategies named feature extraction, fusion and reconstruction to preserve most of the details from the input modalities. We train the parameters of the feature extraction and reconstruction layers by maximising the structural similarity and minimising the euclidean distance between fused image and the input images.

\begin{figure}[!htbp]
\centering
\includegraphics[width=\textwidth]{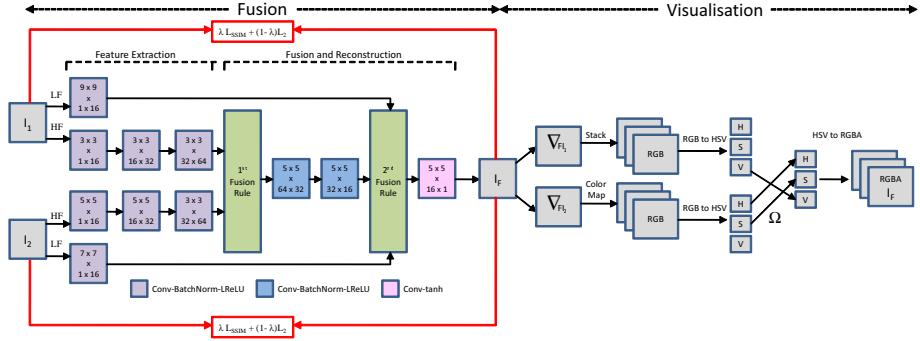}
\caption{The proposed fusion and visualisation framework.} \label{fig1}
\end{figure}

\subsubsection{Feature Extraction:} In the first feature extraction layer, we perform two different convolution operations on each of the input images to decompose it into high and low frequency feature maps. Since blurry PET images has higher low frequency components than sharp MRI images, we define a kernel filter of size $9\times9$ for the anatomical input $I_{1}$ to capture low frequency (LF) features in a larger window while we select a smaller kernel size of $7\times7$ to capture the LF features of the functional input $I_{2}$ efficiently. For the high frequency (HF) layer, we define a kernel size of $3\times3$ for anatomical input $I_{1}$ to capture the sharp local features such as edges and corners better in smaller neighborhoods while we choose a kernel size of $5\times5$ for functional input $I_{2}$ due to less number of edges. We add two more hidden HF layers with increasing number of channels to preserve the deep high frequency features at the boundary regions.
\subsubsection{Fusion and Reconstruction:}
HF features contain detailed information about texture and edges that has direct impact on the edge distortion of the fused image. Therefore, proper selection of the fusion strategy of HF features is crucial for robust fusion results. Max pooling strategy extracts edges from the features maps whereas average pooling is efficient in preserving textures. We utilise the advantage of each of the methods and propose max-average pooling as $1^{st}$ fusion rule for the HF features. We implemented weighted averaging strategy as the $2^{nd}$ fusion rule for LF features containing global information of inputs. Our reconstruction strategy contains three hidden layers and we define $tanh$ activation function at the last layer due to its steeper
gradients than a sigmoid function making backpropagation effective. Let $H_1(\phi)$ and $H_2(\phi)$ be the high frequency features of $I_1$ and $I_2$ at channel $\phi$ in the third hidden HF layer, $L_1(\tau)$ and $L_2(\tau)$ the low frequency features of $I_1$ and $I_2$ at channel $\tau$ in the first hidden LF layer and $R_{i}(\tau)$ the feature map generated from the second reconstruction layer, then the outputs of first fusion layer $H_{o}(\phi)$ and the second fusion layer $R_{o}(\tau)$ are:
\begin{equation}
H_o(\phi) = \frac{max\Bigl(H_1(\phi), H_2(\phi)\Bigl)} {H_1(\phi) + H_2(\phi) }, \mkern9mu
R_{o}(\tau) = \frac{L_{1}(\tau) + L_{2}(\tau) + R_{i}(\tau)}{3}
\end{equation}


\subsubsection{Loss function:}
The
fused image in medical domain is normally evaluated by a
human observer whose sensitivity to noise depends on local
luminance, contrast and structural properties of the image. Therefore, we adopt the structural similarity
index ($SSIM$ \cite{ref19}) as the human perceptive loss
function defined as:
\begin{equation}
SSIM(I, J) = \frac{1}{N} \sum_{k=1}^{N} [l(i_k,j_k)]^{\alpha}\cdot [c(i_k,j_k)]^{\beta}\cdot [s(i_k,j_k)]^{\gamma}
\end{equation}
where $I$ and $J$ are the two input images and $N$ is the number of local
windows in the image. In our paper, $\alpha$=$\beta$=$\gamma$=1 gives equal importance
to luminance $l(i_k , j_k)$, structural $s(i_k , j_k)$ and contrast $c(i_k , j_k)$ comparisons of the image
contents $i_k$ and $j_k$ at $k^{th}$ local window with $C_l$, $C_c$, $C_s$ as constants given as:
\begin{equation}
l(i_k, j_k) = \frac{2\mu_{i_k}\mu_{j_k}+ C_l}{\mu_{i_k}^{2}\mu_{j_k}^{2}+C_l}, \mkern9mu
c(i_k,j_k) = \frac{2\sigma_{i_k}\sigma_{j_k} + C_c}{\sigma_{i_k}^{2}\sigma_{j_k}^{2} + C_c},
\mkern9mu s(i_k,j_k) = \frac{\sigma_{i_k j_k} + C_s}{\sigma_{i_k} + \sigma_{j_k} + C_s}
\end{equation}
where $\mu_{i_k}$, $\mu_{j_k}$ are the mean and $\sigma_{i_k}$, $\sigma_{j_k}$
are the standard deviations of image contents $i_k$
and $j_k$ computed using a Gaussian filter with standard deviation $\sigma_g$ and $\sigma_{i_k j_k}$ being the correlation coefficient. By empirically setting only SSIM as the loss function, we observed a shift in brightness of the fused image since the smaller $\sigma_g$ preserves edges and contrast better than the luminance in the flat areas of the image. Therefore, in addition to SSIM we employ pixel level loss $\ell_2$ which preserves luminance better. With $I_1$ and $I_2$ as the two source images and $F$ as the final fused image, we express our steerable total
loss function as:
\begin{equation}
L_{total} = \lambda * L_{SSIM} + (1- \lambda) * L_{\ell_{2}}
\end{equation}
where $L_{SSIM} = (1 - SSIM(I_1, F)) + (1 - SSIM(I_2, F))$ and $L_{\ell_{2}} = (||F - I_1||_2 + ||F - I_2||_2)$ while $\lambda$ controls the weightage of each of the sub-losses.

\subsection{Visualization framework}

We visualised the functional and anatomical information in the fused grayscale image by first calculating the partial derivative of each pixel of the fused image with respect to each of the input images. Assuming $n$ and $m$ as the dimensions of the anatomical input $I_1$ and functional input $I_2$ while $k$ and $l$ are the dimensions of the fused image $I_F$, so the gradients $\nabla_{FI_{1}}(n,m)$ and $\nabla_{FI_{2}}(n,m)$ will be:

\begin{equation}
\nabla_{FI_{1}}(n,m) = \sum_{i=0}^{k}\sum_{j=0}^{l} \frac{\partial F[i,j]}{\partial I_{1}[n,m]}, \mkern9mu
\nabla_{FI_{2}}(n,m) = \sum_{i=0}^{k}\sum_{j=0}^{l} \frac{\partial F[i,j]}{\partial I_{2}[n,m]}
\end{equation}


We then color coded the functional gradient image $\nabla_{FI_{2}}$ and performed Hue Saturation Value (HSV) transformation on both the images. The Hue and Saturation channels of $\nabla_{FI_{2}}$ and the Value channel from $\nabla_{FI_{1}}$ were stacked and inverse transformed to get the fused colored image. The factor $\Omega$ is multiplied with the saturation channel of $\nabla_{FI_{2}}$ to prevent the occlusion of anatomical information. 

\begin{figure}
\centering
\includegraphics[width=0.9\textwidth]{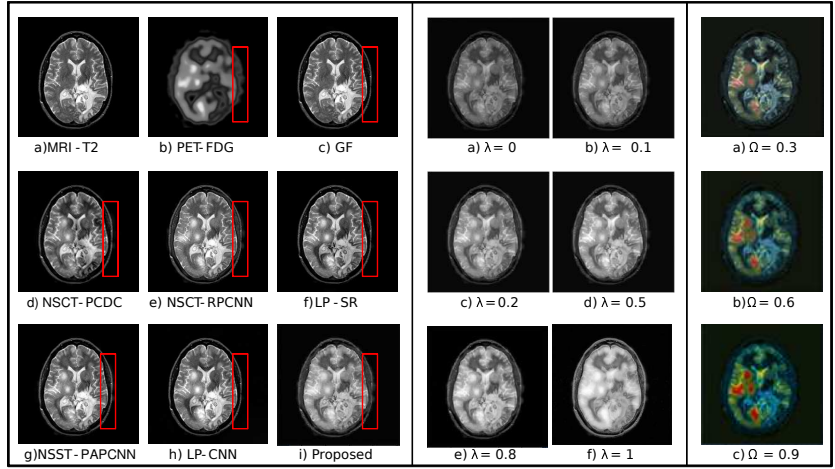}
\caption{The three sets of images shows visual results of compared methods, proposed fusion results based on $\lambda$ and the visualisation results based on $\Omega$.} 
\label{fig2}
\end{figure}

\section{Experiments and results}
\subsection{Training}
\subsubsection{Data acquisition:}
We obtained 500 MRI-PET image pairs publicly available at the Alzheimer's Disease Neuroimaging
Initiative (ADNI) \cite{ref20} with subject's age between 55-90 years among both genders. All images were analyzed as axial slices with a voxel size of 1.0 x 1.0 x 1.0 $mm^{3}$. The MRI images were skull stripped T1 weighted N3m MPRAGE sequences while PET-FDG images were co-registered, averaged, standardized voxel sized with uniform resolution of the same subject. We aligned the MRI-PET image pairs using the Affine transformation tool of 3D Slicer registration library.
\subsubsection{Initialisation of hyperparameters:} 
The kernel filters of our fusion network are initialised as truncated normal distributions with standard deviation of 0.01 while the bias is zero. The stride in each layer is 1 with no padding during convolution since every down-sampling layer will erase detailed information in the input images which is crucial for medical image fusion. We employ batch normalization and Leaky ReLU activation with slope 0.2 to avoid the issue of vanishing gradient. The network is trained for 200 epochs with the batch size of 1 and varied $\lambda$ $\in$ [0,1] on a single GeForce GTX 1080 Ti GPU. The Adam optimizer is
used as the optimization function during backpropagation step
with learning rate of 0.002. Our approach has been implemented in Python
2.7 and Tensorflow 1.10.1 on a Linux Ubuntu 17.10 x86\_64
system with 12 Intel Core i7-8700K CPU @ 3.70 GHz and
64-GB RAM. 

\begin{figure}[!htb]
\begin{subfigure}{.333\textwidth}
  \includegraphics[width=\linewidth]{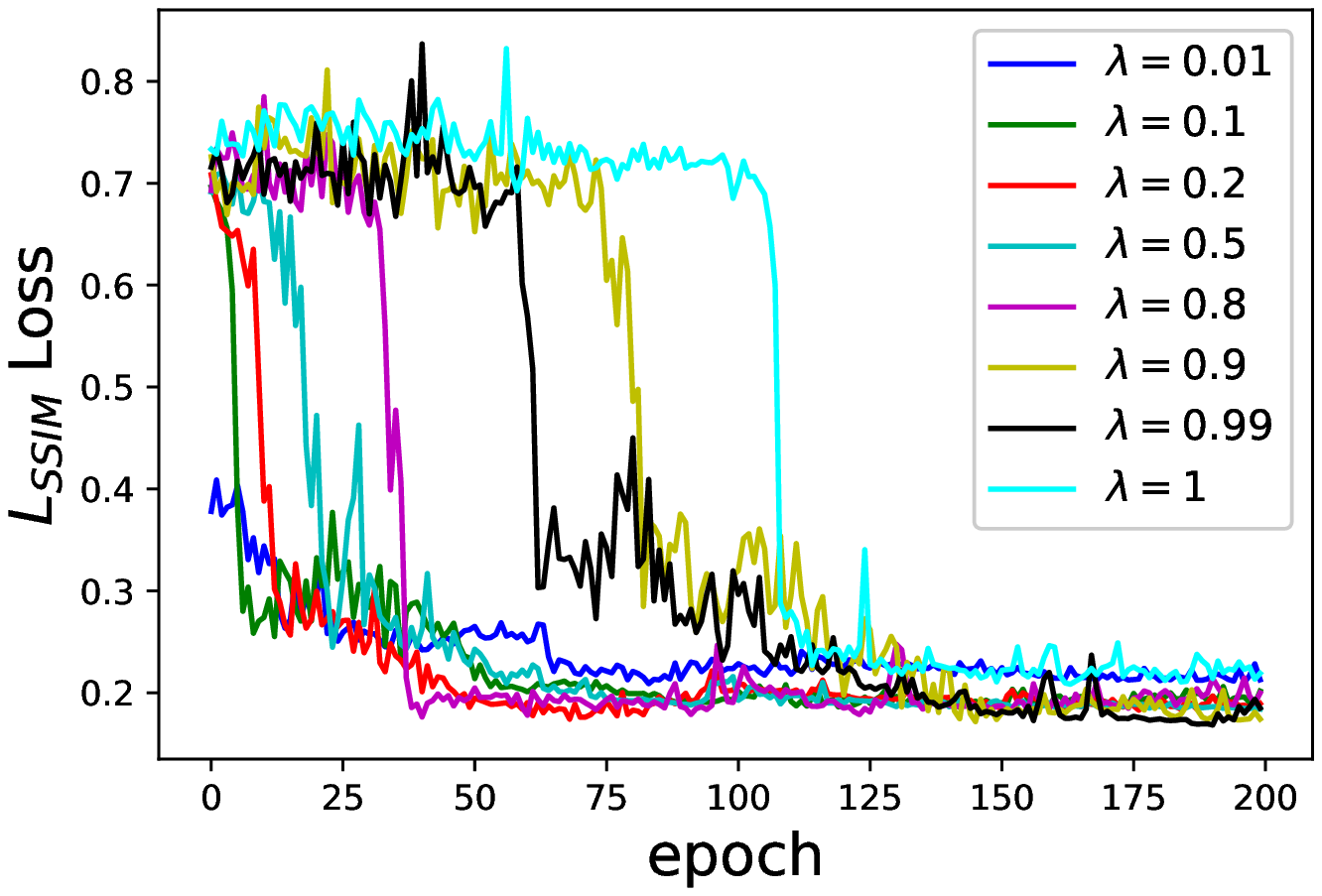}
  \caption{MRI $L_{SSIM}$ Loss}
\end{subfigure}%
\begin{subfigure}{.333\textwidth}
  \includegraphics[width=\linewidth]{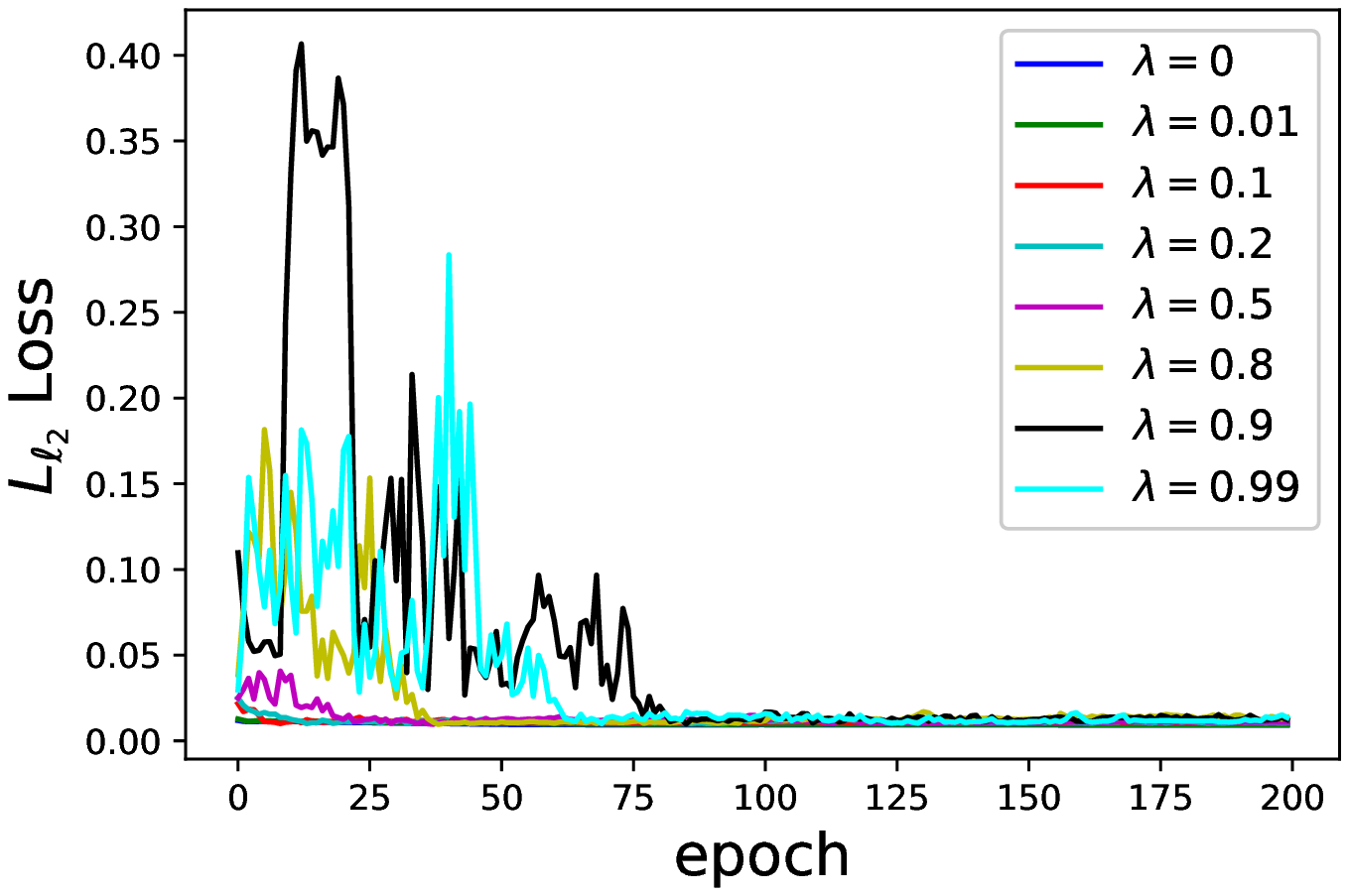}
  \caption{ MRI + PET $L_{\ell_2}$ Loss}
\end{subfigure}%
\begin{subfigure}{.333\textwidth}
  \includegraphics[width=\linewidth]{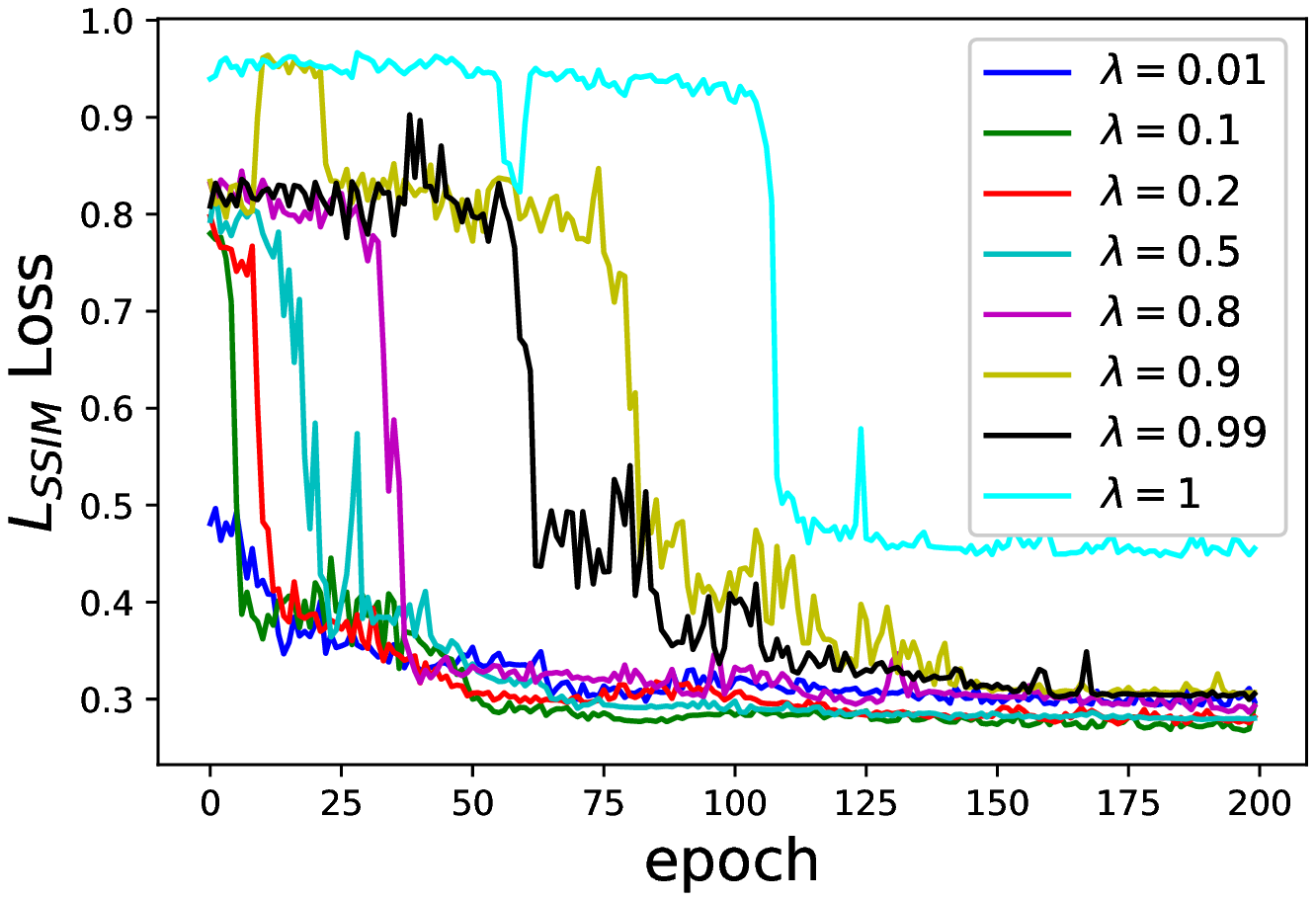}
  \caption{ PET $L_{SSIM}$ Loss}
\end{subfigure}%
\caption{Training loss curves with 200 epochs and several $\lambda's$.}
\end{figure}

\subsubsection{Loss curve analysis:} 
Fig.3. shows the loss curves $L_{SSIM}$ and $L_{\ell_2}$ for the training data at different values of $\lambda$. The figures convey rapid convergence for all $\lambda$ values other than $\lambda \geq 0.9$ where $L_{SSIM}$ plays more important role than $L_{\ell_2}$ in the total loss function $L_{total}$. It is to be noted that $L_{SSIM}$ has higher sensitivity to smaller errors such as luminance variations in flat texture-less regions while $L_{\ell_2}$ is more sensitive to larger errors irrespective of the underlying regions within the image. This property leads to delayed convergence of $L_{SSIM}$ for visually perceptive results at edges as well as flat regions of the fused image. 

\subsection{Testing}

We performed cross-validation on our trained model with a disjoint test dataset that contain 100 MRI-PET image pairs of 100 unique subjects from ADNI and Harvard Whole Brain Atlas \cite{ref21} databases. 90 MR-T1 and PET-FDG image pairs obtained from ADNI were mutually exclusive from training image pairs. In order to test our method on datasets distinct from ADNI, the remaining 10 pre-registered image pairs were a combination of MR-T1 and PET-FDG or MR-T2 and PET-FDG images obtained from Harvard Whole Brain Atlas \cite{ref21} with subjects suffering from either Glioma or Alzeihmer's disease.

\subsection{Evaluation settings}

The visualisation results of the test images were evaluated with 10 values of $\lambda, \Omega \in [0,1]$
on four objective assessment metrics namely nonlinear correlation information
entropy ($Q_{IE}$) \cite{ref22}, xydeas metric ($Q_{G}$) \cite{ref23}, feature mutual information ($Q_{FMI}$) \cite{ref24},
structural similarity metric ($Q_{SSIM}$) \cite{ref19} and human perceptive
visual information fidelity ($Q_{VIFF}$) \cite{ref26} with higher values means better performances. The evaluation resulted in highest scores with $\lambda$ = 0.8 and $\Omega$ = 0.6 for three of the mentioned metrics. We then used six different medical image fusion
methods from recent past namely guided filtering
(GF) \cite{ref6}, nonsubsampled contourlet transform (NSCT-PCDC)
\cite{ref2} and (NSCT-RPCNN) \cite{ref27}, combination of multi-scale transform and
sparse representation (LP-SR) \cite{ref9}, nonsubsampled shearlet
transform (NSST-PAPCNN) \cite{ref5} and convolutional neural
networks (LP-CNN) \cite{ref_proc15} for quantitative comparisons in a MATLAB R2018a environment. Our code is publicly available at: \url{https://github.com/nish03/FunFuseAn/}.

\begin{table}
\centering
\caption{Assessment of fusion methods based on objective metrics and runtime.}\label{tab1}
\begin{tabular}{|p{1.2cm}|p{1.2cm}|p{1.1cm}|p{1.2cm}|p{1.3cm}|p{1.4cm}|p{1.1cm}|p{1.3cm}|}
\hline Metrics & GF & NSCT-PCDC & LP-SR & NSCT-RPCNN & NSST-PAPCNN & LP-CNN & Proposed \\
\hline
$Q_{IE}$ &  0.8169 & 0.8080  & 0.8092 & 0.8132 & 0.8102  & 0.8076 & 0.8104\\
$Q_{G}$ &  0.7555  & 0.5457  & 0.6501 & 0.6702 & 0.6685 & 0.5665 & 0.5707\\
$Q_{FMI}$ & 0.9224 & 0.8754  & 0.8969 & 0.8941 & 0.8997 & 0.8958 & 0.8885\\
$Q_{SSIM}$ & 0.8260 & 0.7992 & 0.7837 & 0.8492 & 0.8318 & 0.7176 & 0.8610\\
$Q_{VIFF}$ & 0.2776 & 0.3415 & 0.5990 & 0.5430 & 0.6001   & 0.5326 & 0.6005\\
\hline
Time (s) & 13.43 & 221.07 & 75.69 & 775.31 & 521.36 & 481.73 & 0.37\\ 
\hline
\end{tabular}
\end{table}

\subsection{Comparison to the state of the art}
\subsubsection{Visual results:} The first set of Fig.2. conveys negligible contribution of PET features in the fused image by GF while NSCT-PCDC, NSST-PAPCNN, LP-SR and NSCT-RPCNN has
uneven distribution of structural edges and contrast leading to splotchy visual
artifacts. The results from LP-CNN are better than other
methods but like other methods it fails to preserve the edges from
functional modality i.e. PET. Our method conserve
structural information better in both of the image pairs and is robust
in preserving the edges (see PET features in red box). The second set of Fig.2. reveals that the luminance of the proposed fusion results increases with greater $\lambda$ values leading to brightness artifacts at corner cases of $\lambda$ = 0 and $\lambda$ = 1. The third set of Fig.2. shows proposed visualisation results at $\lambda=0.8$ controlled by parameter $\Omega$ where a shift in occlusion of the anatomical information with different values of $\Omega$ could be observed.


\subsubsection{Objective assessment:}
Table 1. summarizes the average scores of 100 test image pairs computed for different fusion methods along with our proposed method at $\lambda$ = 0.8 and $\Omega$ = 0.6. A method with a higher score performs
better than a method with a lower score which is applicable for all the mentioned metrics. The results convey that our method performs better with the quality metric
$Q_{SSIM}$ and $Q_{VIFF}$. This is assertive from the fact that the neural network optimizes the loss function
and subsequently improves the structural information
in the fused image. Overall, the competitive scores reflects the robustness of our method for human
perceptive fusion results.

\subsubsection{Computational Efficiency:}
We evaluated the total runtime of each of the methods for 100 test images in the MATLAB R2018a
environment. 
Table 1. conveys that our fusion and visualisation method achieved
best timings since the network parameters are optimized during the
training phase and with a fixed batch size it requires just one forward propagation through the fusion network to generate the
fused images. Therefore, our fusion network could also be utilized in
a real time neurosurgical intervention setup where a continuous feed of live images in a form of time series 
will generate fused output video stream with very low time delay.

\section{Conclusion and Discussion}
We presented a novel image fusion and visualisation framework which is highly suitable for diagnosing malignant brain conditions. The end-to-end learning based fusion model utilised the structural similarity loss to construct artifact free fusion images and the gradient based visualisation delineated the anatomical features of MRI from the functional features of PET in the fused image. The extensive evaluation of our approach conveyed significant improvements in human perceptive results compared to past methods. In future, our method could further be extended to include other combination of anatomical and functional imaging modalities by changing the fusion architecture especially the feature extraction layers. Additionally, we plan to immersively visualise the proposed results in an augmented reality based real time preoperative setup, thereby enabling medical experts to make robust clinical decisions.  

\subsubsection{Acknowledgements.}
This work was supported by the European Social Fund (project no. 100312752) and the Saxonian Ministry of Science and Art.

\end{document}